# Performance Metrics and Loss Mechanisms in Horticulture Luminescent Solar Concentrators

*Zhijie Xu[1], Yue Yu[1,2], Ioannis Papakonstantinou [1] **

Zhijie Xu, Yue Yu, Prof. Ioannis Papakonstantinou
Photonic Innovations Lab, Department of Electronic and Electrical Engineering
University College London
London WC1E 7JE, UK

Yue Yu
School of Optoelectronic Engineering
Xidian University
Xi'an 710071, Shaanxi, China

E-mail: i.papakonstantinou@ucl.ac.uk

**Abstract**

Horticulture Luminescent Solar Concentrators (HLSCs) represent an innovative concept developed in recent years to promote crop yields, building upon the foundation of traditional Luminescent Solar Concentrators (LSCs). HLSCs are characterized by two distinct properties: spectral conversion and light extraction. Unlike traditional LSCs, HLSCs focus on converting energy from one part of the solar spectrum (typically green) to a specific range (usually red) and aim for the converted photons to exit the device from the bottom surface rather than the edge surfaces. In this study, we start by examining the specific requirements of horticulture to clarify the motivation for using HLSCs. We re-evaluate and propose new optical metrics tailored to HLSCs. Additionally, we analyse potential loss channels for direct red emission and converted red emission. Utilizing Monte Carlo ray tracing method and experimental data, we further explore the factors influencing these loss channels. Our work provides a fundamental discussion on HLSCs and offers design guidelines for future HLSC research.

**1. Introduction**

It is widely accepted that light plays the most dominant role in plant growth among all environmental parameters[1,2]. This influence can be attributed to three key aspects: light intensity, photoperiod, and spectral distribution[3,4]. Light intensity refers to the amount of light that plants receive, within the Photosynthetically Active Radiation (PAR) range typically between 400-700 nm[5]. However, ultraviolet (UV) and far-red light also play role to plant growth, and so in this study it was decided to extend the spectral range of interest to cover the 300-900 nm band, which we refer to as the Extended PAR, or EPAR, as shown in Figure 2b. The photoperiod pertains to the duration of light exposure that plants receive within a 24-hour period[6]. Spectral distribution describes the intensity of different colours in the radiation spectrum, as various colours exert distinct effects on plant growth[7].

The various effects of light on plant growth are illustrated in Figure 1. Excessive UV light can reduce photosynthesis and potentially cause cell death[8]. Blue light inhibits cell expansion, thereby reducing leaf area[9]. Most plants exhibit weak absorption of green light[10]. In contrast, chlorophyll exhibits the strongest absorption in the red, which tends to stimulate root formation and seed germination[11]. Far-red light aids in flower regulation[12]. It is evident that short-wavelength components (less than 600 nm) generally show lower potential for plant growth compared to long-wavelength components[13].

However, the actual solar spectrum peaks in the green range, leading to suboptimal utilization of solar energy by plants. As depicted in Figure 2b, the relatively inefficient green light (approximately 500 to 600 nm) constitutes 21.5% of the AM 1.5 solar power within the EPAR spectrum, representing the highest percentage compared to other colours. If we consider only the PAR spectrum, this percentage rises to 35.7%. This mismatch between plant light preferences and solar spectrum distribution highlights an innovative solution to enhance energy utilization: spectral conversion[14–16]. Effective spectral conversion schemes for plant growth may include both down-conversion (green to red) and up-conversion (near-infrared to red)[14–19]. In this paper, we primarily focus on down-conversion, as up-converting fluorophores have not yet reached the same level of efficiency [20].

Luminescent solar concentrators, which consist of two components, host matrix and fluorescent inclusions, can be efficient tools for realizing spectral conversion[21–24]. However, conventional LSCs primarily focus on trapping photons within the host matrix hence, the "concentrator" part in their name. This high-efficiency trapping is attributed to Total Internal Reflection (TIR)[25]. Given that the most commonly used host materials in LSCs are polymers and glasses with refractive indices around 1.5, over 70% of the converted red photons are typically trapped within the device and eventually guided to the edges[26]. In contrast, for a HLSC device, the goal is to allow the converted red photons to pass through the host matrix and be received by the crops, as shown in Figure 2c.

To improve the outcoupling efficiency of photons from the bottom surface (the exit side facing the plants), it is effective to apply light extraction techniques. Typical light extraction techniques, such as lowering the refractive index and incorporating micro/nano structures, usually involve modifications to the host matrix[27–30]. Consequently, the efficiency of the converted red emission would be significantly enhanced, but the original red photons in the EPAR spectrum could also be affected. Considering that red light (600-700 nm) accounts for 20.2% of the EPAR spectrum solar power (as shown in Figure 2b), the second highest percentage, we cannot overlook the impact that any modification to the LSC structure may have on the original red photons.

Conventional LSCs inherently reduce the original red emission from the solar spectrum, even without integrating light extraction techniques. This reduction is primarily due to two factors. Firstly, most host polymers absorb a small amount of light, albeit this absorption is usually very small[23,31]. Secondly, the Stokes shift of most fluorescent materials is not ideally large, leading to an overlap between the absorption and emission spectra[21,22]. As a result, many fluorophores that re-emit red photons also absorb red photons. This effect reduces both the number of converted red photons and original red photons.

In conclusion, the usable red photons for crops originate from two sources: converted red photons and direct red photons. Both are crucial to the design of optimum HLSCs. Therefore, the core objective of HLSC design is to maximize the total number of red photons reaching the plants, which rationally involves extracting more converted red photons while preserving the direct red photons. Here, we begin with an analysis of the loss channels for both direct and converted red emission in Sections 2.1 and 2.2, respectively. The metrics used in HLSCs are defined and discussed in Section 2.3. The most significant factors affecting these loss channels are analysed in detail using the Monte Carlo ray tracing method and experimental results, which are covered from Sections 3.1 to 3.5. Potential optimization methods are explored for future research in Section 4.

## 2. Challenges and Background

Before discussing the loss channels of HLSCs, we have made the following assumptions to facilitate understanding. In real applications, HLSC devices can have various shapes, such as circular disks, rectangular or polygonal slabs. Regardless of their shape, we classify all the surfaces of HLSC devices into three categories: top surface, bottom surface, and edge surfaces, as shown in Figure S1. The top surface is the surface the HLSC receives sunlight from. The bottom surface is the surface opposite the top surface, positioned close to the plants. The edge surfaces are the surfaces surrounding the top and bottom surfaces, usually perpendicular to them.

### 2.1. Loss channels of direct red photons

The main loss channels of direct red photons are illustrated in Figure 3b. Firstly, the top surface of the HLSC back-reflects a portion of these photons. This back-reflection is due to Fresnel reflections and influenced by the incident angle of sunlight. Even with normal incidence, over 4% of the incident red light is back reflected (for n~1.5)[32]. As the sun moves during the day, the incident angle increases significantly. When the incident angle exceeds 70°, the reflection loss can increase to approximately 17%. The effect of angle variation on reflection loss is shown in Figure S2.

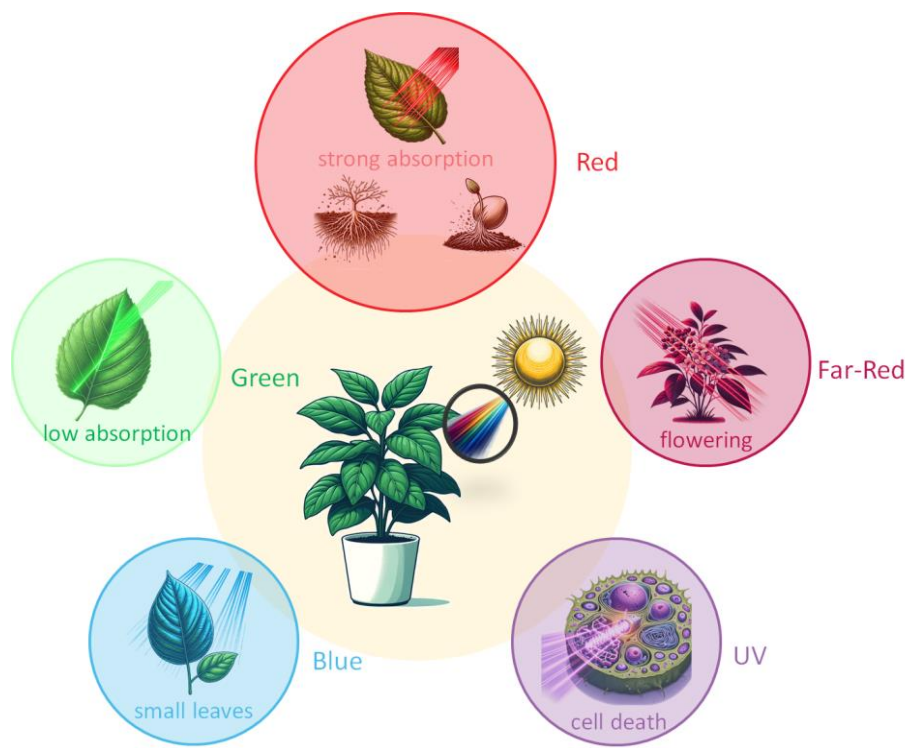

**Figure 1.** The effects of different colours in the solar spectrum on plant growth and development.

Similarly, red photons are also back-reflected by the bottom surface. Additionally, inhomogeneities within the host matrix can introduce a small amount of scattering, as was observed previously[33]. The presence of fluorophores in the HLSC may also scatter the direct red photons. Both sources of scattering result in a wider angular distribution of light hitting the bottom surface, occasionally increasing the incident angle beyond the critical angle, preventing some light from escaping.

Thirdly, due to limitations on the magnitude of Stokes shift, some absorption of direct red photons is almost unavoidable. This absorption leads to two loss channels. First, due to the non-unity quantum yield (QY), some absorbed photons are not re-emitted and are lost as heat. Second, even if photons are re-emitted by the fluorophores, it is still not guaranteed that they will escape the HLSC (see loss channels of converted photons below).

**2.2. Loss channels of converted red photons**

The loss channels of converted red photons are more complex due to the photoluminescence process, as depicted in Figure 3c. Firstly, the absorption spectrum of the fluorescent materials typically cannot cover the entire range of unused solar spectrum, leading to low spectral conversion efficiency. Additionally, due to the non-unity QY, some absorbed photons are lost as heat rather than being re-emitted as converted red photons. Even if the converted photons are successfully emitted by the fluorophores, it remains challenging for them to exit the device through the bottom surface. Given that in most typical HLSCs re-emission is isotropic, the emission directions of converted red photons are evenly distributed within a $4\pi$ steradian range[34]. Due to TIR, ~70% of the re-emitted converted red photons are trapped within the device.

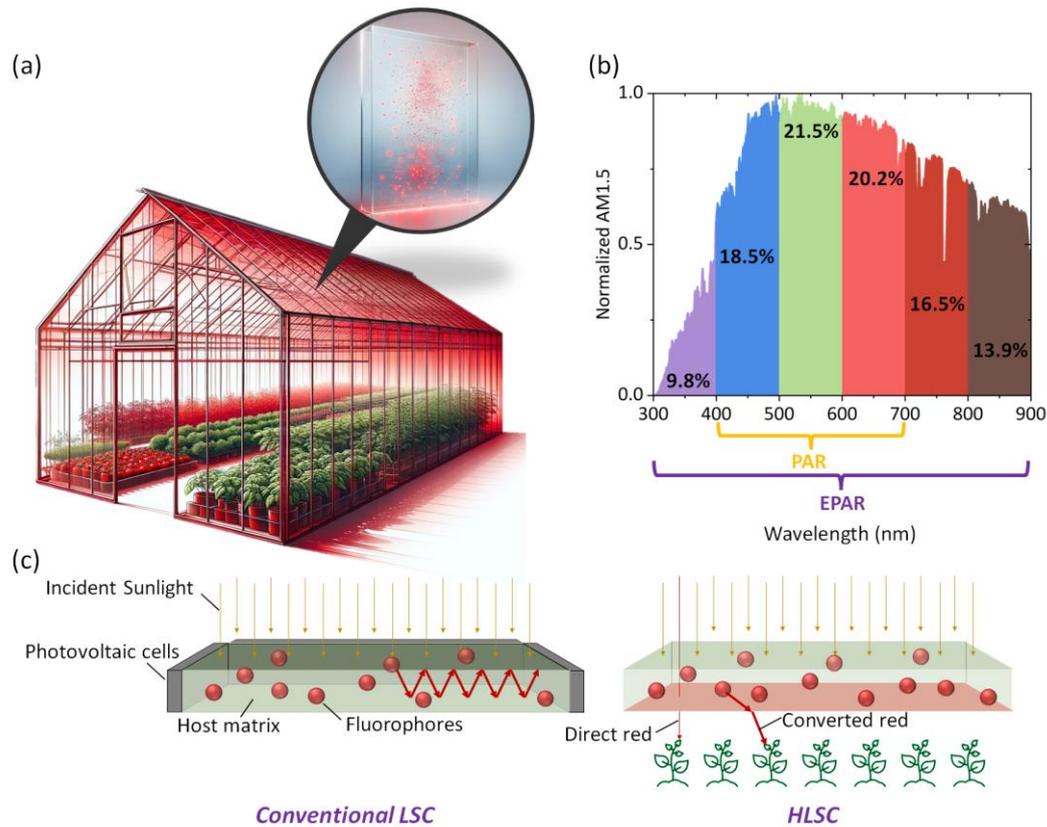

**Figure 2.** a) Illustration of a greenhouse covered with HLSC. The enlarged inset shows a transparent host matrix doped with red fluorophores. b) Normalized AM1.5 spectrum within the range of the EPAR spectrum. The numbers shown in each spectral range indicate the weight ratio of the respective colours. We define the spectral ranges as follows: 300-400 nm as UV, 400-500 nm as blue, 500-600 nm as green, 600-700 nm as red, 700-800 nm as far-red, and 800-900 nm as near-infrared. c) Comparison between the conventional LSC and HLSC. Conventional LSCs trap light and direct it to the edges, while the objective of HLSCs is to direct converted red photons to the plants from the bottom surface.

Commonly, to maximize the absorption of non-useful photons, fluorophores are typically doped at high concentrations in HLSCs, which increases the likelihood of reabsorption. Reabsorption raises the probability of converted red photons being lost due to QY loss and being back-reflected from the bottom surface.

**2.3. Metrics in HLSC**

Building upon the aforementioned analysis of loss channels in HLSCs, new metrics should be discussed and defined to better evaluate the performance of these devices. It is clear that the ultimate goal of HLSC design is to improve the red emission from the bottom surface, which we will refer to as Total Red Emission (TRE, $\eta_{TRE}$). In detail, $\eta_{TRE}$ can be expressed as:

$$\eta_{TRE} = \frac{n_{\text{total red}}}{n_{\text{input}}}, (1)$$

where $n_{\text{total red}}$ is the number of total red photons emitted from the bottom surface and $n_{\text{input}}$ represents the total number of photons from the incident spectrum within the EPAR spectrum. Detailed illustration of these quantities is provided in Figure S3. In order to better understand $\eta_{TRE}$, this metric is divided into Direct Red Emission (DRE, $\eta_{DRE}$) and Converted Red Emission (CRE, $\eta_{CRE}$). These two metrics can be expressed as:

$$\eta_{DRE} = \frac{n_{\text{direct red}}}{n_{\text{input red}}}, (2)$$

$$\eta_{CRE} = \frac{n_{\text{converted red}}}{n_{\text{input red free}}}, (3)$$

where $n_{\text{input red}}$ is the number of total photons from the incident spectrum that includes only red components (600-700 nm), and $n_{\text{input red free}}$ is the number of total photons from the incident spectrum excluding the red components. $n_{\text{direct red}}$ represents the number of direct red photons escaping from the bottom surface, and $n_{\text{converted red}}$ represents the number of converted red photons escaping from the bottom surface correspondingly. It is important to clarify that $\eta_{DRE}$ also includes red photons that are absorbed by fluorophores, re-emitted, and then exit the HLSC. This quantity accounts for such secondary (or even tertiary, quaternary, etc.) effects.

The relationship between $\eta_{CRE}$, $\eta_{DRE}$, and $\eta_{TRE}$ depends on the percentage of red photons present in the incident solar spectrum. Let $\rho_d$ representing the weight of red in EPAR, and $\rho_c$ representing the weight of the remaining components. Their relationship can be expressed as:

$$n_{\text{input}} = n_{\text{input red}} + n_{\text{input red free}}, (4)$$

$$n_{\text{total red}} = n_{\text{direct red}} + n_{\text{converted red}}, (5)$$

$$\eta_{\text{TRE}} = \eta_{\text{DRE}} \times \rho_d + \eta_{\text{CRE}} \times \rho_c, (6)$$

where $\rho_d + \rho_c = 1. (7)$

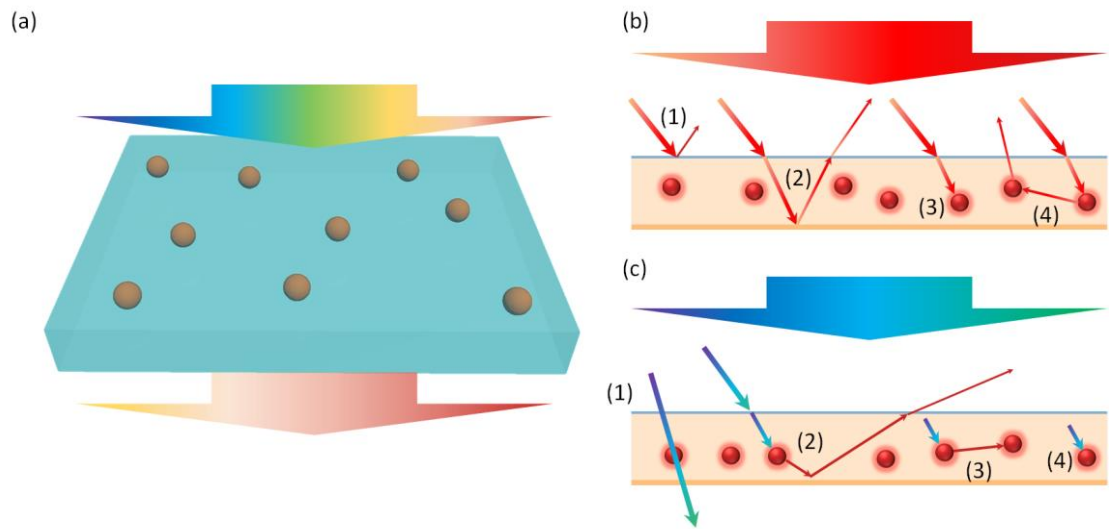

**Figure 3.** a) HLSCs optimize the spectral distribution of incident solar spectrum to enhance plant growth. b) Potential loss channels for direct red photons (from left to right): 1) Fresnel reflection; 2) back reflection from the bottom surface; 3) QY loss; 4) reabsorption loss. c) Potential loss channels for converted red photons (from left to right): 1) not absorbed; 2) back reflection from the bottom surface; 3) reabsorption loss; 4) QY loss. In Figures b) and c), the blue lines represent the top surface, while the red-orange lines represent the bottom surface, as shown in Figure S1.

Several parameters influence the value of $\eta_{CRE}$ and $\eta_{DRE}$. Fresnel loss ($F_l$) represents the ratio of photons that are back reflected by the top surface of the HLSC:

$$F_l = \frac{n_{input} - n_{incoming}}{n_{input}}, (8)$$

where $n_{incoming}$ is the number of photons left after the back reflected ones have been excluded. The potential fates of incoming photons are: 1) absorbed by host materials, 2) absorbed by fluorophores, and 3) unabsorbed, as shown in Figure S3. Thus, $n_{incoming}$ could be expressed as:

$$n_{incoming} = n_{absorbed} + n_{host} + n_{unabsorbed}, (9)$$

where $n_{host}$ represents the number of photons absorbed by the host material. $n_{absorbed}$ represents the number of photons absorbed by the fluorophores and $n_{unabsorbed}$ represents the number of photons that are not absorbed at all. Based on the above metrics, the internal absorption ($A$) in LSCs is usually expressed as:

$$A(\lambda) = \frac{n_{absorbed}}{n_{incoming}}. (10)$$

However, in HLSC, external absorption ($A_{ex}$) is a more accurate metric for overall device performance evaluation, defined as:

$$A_{ex}(\lambda) = \frac{n_{absorbed}}{n_{input}}, (11)$$

where $\lambda$ indicates that the absorption for red photons and red-free light is different. When $\lambda$ = red, $A$ and $A_\text{ex}$ refer to absorption of red photons. When $\lambda = \overline{\text{red}}$, they refer to absorption for red-free light (i.e. the rest of the photons in EPAR excluding 600-700 nm). For direct red emission, $n_\text{direct red}$ can be expressed as:

$$n_\text{direct red} = n_\text{unabsorbed}(\text{red}) \times T_b + n_\text{absorbed}(\text{red}) \times QY \times E_s \text{ , (12)}$$

where $T_b$ represents the transmittance of the bottom surface at the specific angle of incidence of direct red photons, and $E_s$ represents the escape efficiency of the device, corresponding to the ratio of *isotropically* re-emitted red photons that successfully leave the bottom surface. Here, we assume that the emission for both spherical quantum dots and randomly oriented dipoles (such as dyes) is isotropic, as has been shown previously[35].

Similarly, for the converted red emission, $n_\text{converted red}$ could be expressed as:

$$n_\text{converted red} = n_\text{absorbed}(\overline{\text{red}}) \times QY \times E_s. \text{ (13)}$$

Based on these discussions, internal quantum efficiency (IQE) and external quantum efficiency (EQE) for direct and converted red emission are described as[36,37]:

$$IQE_\text{converted red} = \frac{n_\text{converted red}}{n_\text{absorbed}}, \text{ (14)}$$

$$EQE_\text{direct red} = \frac{n_\text{direct red}}{n_\text{input red}} = \eta_\text{DRE}, \text{ (15)}$$

$$EQE_\text{converted red} = \frac{n_\text{converted red}}{n_\text{input red free}} = \eta_\text{CRE}. \text{ (16)}$$

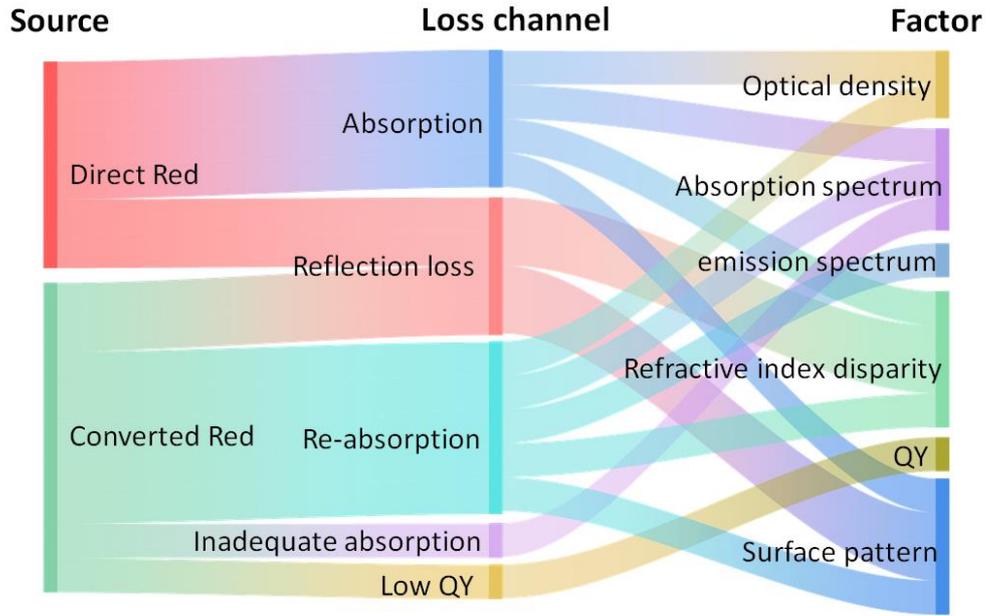

**Figure 4.** Schematic diagram illustrating the source of red photons and the associated potential loss channels. The left bars indicate that the final red photons originate from both direct red photons and converted red photons. The middle bars depict the potential loss channels for both direct red photons and converted red photons. The right bars identify the factors influencing these loss channels.

Table 1 shows the descriptions and expressions for these metrics.

**Table 1.** Metrics of HLSC

| Metric | Description | Expression |
|---|---|---|
| $\eta_{TRE}$ | Ratio of red photons from bottom surface to all input photon numbers | $\eta_{TRE} = \dfrac{n_{\text{total red}}}{n_{\text{input}}}$ |
| $\eta_{DRE}$ ($EQE_{\text{direct red}}$) | Ratio of red photons from bottom surface to input red photons | $\eta_{DRE} = \dfrac{n_{\text{direct red}}}{n_{\text{input red}}}$ |
| $\eta_{CRE}$ ($EQE_{\text{converted red}}$) | Ratio of red photons from bottom surface to input red-free photons | $\eta_{CRE} = \dfrac{n_{\text{converted red}}}{n_{\text{input red free}}}$ |
| Absorption ($A$) | Ratio of photons absorbed by fluorophores to photons entering HLSC | $A = \dfrac{n_{\text{absorbed}}}{n_{\text{incoming}}}$ |

| $IQE_{\text{converted red}}$ | Ratio of red photons from bottom surface to photons absorbed by fluorophores | $IQE_{\text{converted red}} = \dfrac{n_{\text{converted red}}}{n_{\text{absorbed}}}$ |

## 3. Impact of different factors

Combining Equation (1,11-13), $\eta_{\text{TRE}}$ could be expressed as:

$$\eta_{\text{TRE}} = \frac{n_{\text{unabsorbed}}(\text{red}) \times T_b + n_{\text{absorbed}}(\text{red}) \times QY \times E_s}{n_{\text{input}}} + \frac{n_{\text{absorbed}}(\overline{\text{red}}) \times QY \times E_s}{n_{\text{input}}} . \quad (17)$$

As has been reported before, host loss is minimal and can be ignored[38]. This is especially applicable to HLSCs, where the optical path that photons traverse within the device is typically shorter than in traditional LSCs. Proof of this statement is obtained using the Monte Carlo ray tracing method. The mean optical paths in a conventional LSC and an optimized HLSC with a textured bottom surface are compared, as depicted in Figure S4. It is evident that most photons traveling in a conventional LSC are guided toward the edges, resulting in optical path lengths that are several times the lateral size of the device. In contrast, most photons traveling in an HLSC exit from the bottom surface, making their optical path length close to the thickness of the device. Ignoring host absorption and combining Equations (8-11, 17), $\eta_{\text{TRE}}$ can be finally simplified as:

$$\eta_{\text{TRE}} = [1 - A_{\text{ex}}(\text{red}) - F_1] \times T_b + [A_{\text{ex}}(\text{red}) + A_{\text{ex}}(\overline{\text{red}})] \times QY \times E_s. \quad (18)$$

According to Equation (18), it is easy to derive the factors that affect $\eta_{\text{TRE}}$: 1) External absorption ($A_{\text{ex}}$), 2) quantum yield (QY), 3) transmittance of the bottom surface ($T_b$), and 4) escape efficiency of the device ($E_s$). The above parameters can be directly associated to the following physical LSC properties that can be systematically studied and optimized: 1) host-air refractive index contrast; 2) surface patterning; 3) optical density; 4) absorption spectrum; 5) emission spectrum; 6) fluorophore QY. The relationship between loss channels and LSC physical properties for both direct and converted red photons is depicted in Figure 4.

In this paper, several proof-of-concept HLSC devices consisting of Lumogen Red (Sun Chemical Limited) as the fluorophore and PDMS (polydimethylsiloxane, Sylgard 184, Dow Corning) as the host matrix were fabricated and studied to illustrate how the various design parameters affect HLSC performance. Lumogen Red exhibits nearly ideal spectral conversion properties, and PDMS is approximately absorption-free. However, our conclusions are general and applicable to any other host material and fluorophore combination. For comparative purposes, additional experimental results involving Lumogen Yellow, Orange, and Pink (BASF) were also conducted and documented in Figure S9.

### 3.1. Monte Carlo ray tracing

The Monte Carlo ray tracing method is extensively applied in the study of LSCs and is critical for determining optimal parameters in HLSC development too[36–42]. This method allows for continuous adjustment and comprehensive scanning of impacting factors, significantly reducing research time and effort. Additionally, it helps to thoroughly understand underlying mechanisms, enhancing future research efforts.

In real horticultural applications, HLSCs typically have large lateral dimensions that are several orders of magnitude greater than their thickness. In such cases, edge loss can be ignored and HLSCs can virtually be considered as infinitely long. To model such large area devices, we have developed a novel HLSC model with mirror boundaries, inspired by the concept of perfect electromagnetic conducting (PEC) boundaries[43].

Compared to large-area HLSC device simulations, mirror-boundary simulations significantly enhance simulation speed by reducing the size of the simulation domain. For lengths greater than 500 mm, the edge loss is below 0.22%, and for lengths exceeding 5000 mm, the edge loss decreases even further to less than 0.025%. Therefore, in our simulations, the optical properties of the edge surfaces in the HLSC are defined as ideal mirrors (100% reflective), creating an infinite space in the lateral dimensions, as shown in Figure 5. A comparison between our mirror-boundary model and large area LSC, used to prove the accuracy of our method, is detailed in Figure S5. The results indicate that the mirror-boundary HLSC closely resembles real-world scenarios in horticultural applications.

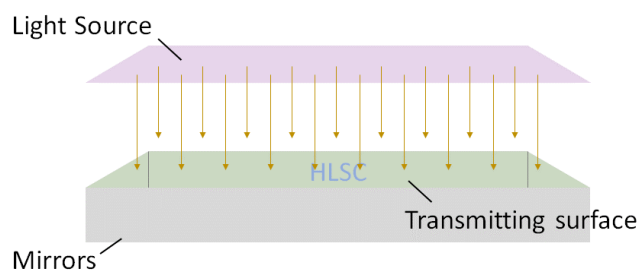

**Figure 5.** Model of HLSC. The purple rectangle represents the light source in the Monte Carlo ray tracing simulation, providing parallel rays. The yellow arrows indicate the rays hitting the HLSC. The bottom slab represents the HLSC. The top and bottom surfaces (green) are transmitting surfaces that share the same optical properties as real HLSC surfaces. The edges (silver) are defined as mirrors to simulate an infinite space in the lateral dimensions.

**3.2. Optical density**

Optical density is a measure of the attenuation of light as it passes through a medium, related to the concentration of absorbing substances and length that light traverses. The expression of optical density (O.D) is given by:

$$O.D = \log_{10} \frac{I_o}{I_t}, \quad (19)$$

where $I_o$ is the intensity of incident light and $I_t$ is the intensity of transmitted light passing through that medium. An example of optical density is illustrated in Figure S6. In the HLSC scenario, optical density is influenced by the device thickness as well as the fluorophore concentration. Mean Free Path (MFP) describes the average distance that photons travel in the host matrix before encountering a fluorophore, effectively reflecting the doping concentration of the fluorophores. Here, a normalized metric defined as thickness/MFP is used to describe the optical density, which consolidates all different concentration/thickness cases into a simpler form. It is evident that the larger the optical density results the larger the absorption ($A_{\text{ex}}(\text{red})$ and $A_{\text{ex}}(\overline{\text{red}})$), leading to an increase in the second term of Equation (18). However, an increase in $A_{\text{ex}}(\text{red})$ would also reduce the first term of Equation (18). Based on this analysis, there exists an optimal optical density that can maximize the $\eta_{\text{TRE}}$ of HLSC.

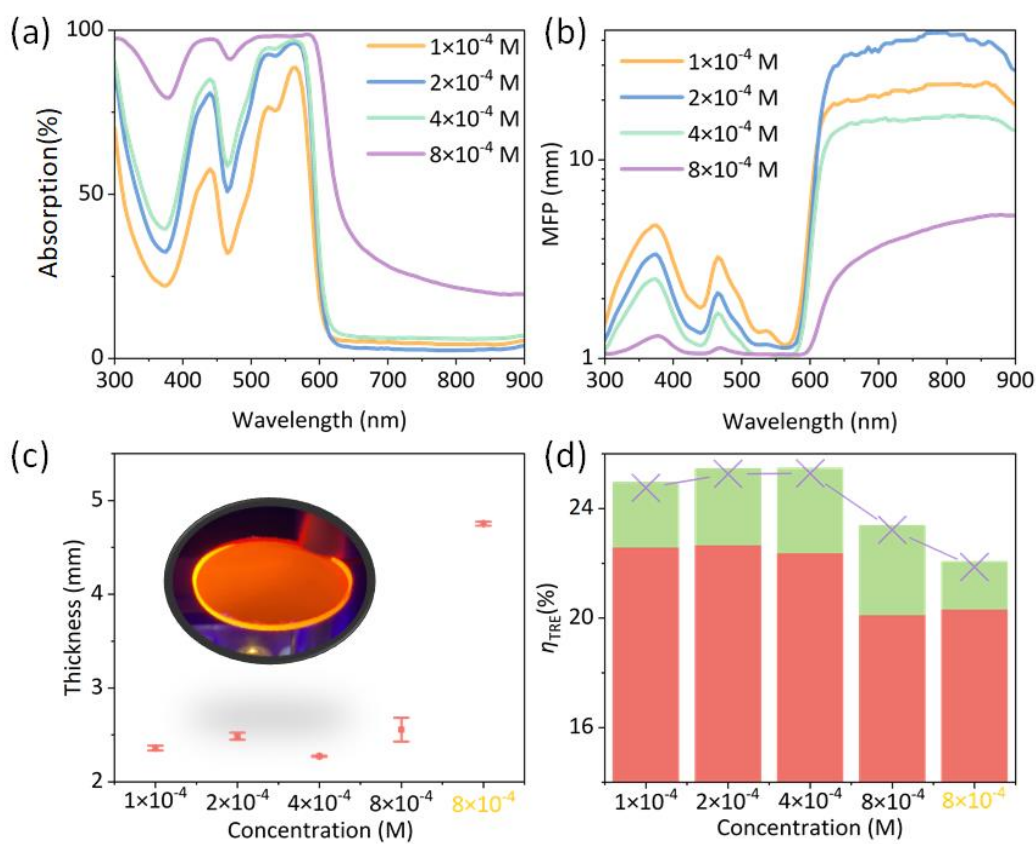

**Figure 6.** a) Measured absorption of HLSC samples with different fluorophore concentrations. b) Calculated MFP of HLSC samples with different fluorophore concentrations. c) Measured sample thicknesses. Inset is a photography of the HLSC with a concentration of $2\times10^{-4}$ M. d) Measured and simulated $\eta_{TRE}$, $\eta_{DRE}$, and $\eta_{CRE}$ results with different optical densities. Green bars and red bars represent the contributions of $\eta_{CRE}$ and $\eta_{DRE}$ to $\eta_{TRE}$, respectively. Purple crosses represent the measured $\eta_{TRE}$ using a solar simulator. Details of measurements conducted with the solar simulator are provided in Figure S7.

Five HLSC samples with different optical densities were made to investigate the effect of optical density on $\eta_{TRE}$. As shown in Figure 6a, dyes were doped in PDMS at four concentrations: $1\times10^{-4}$ M, $2\times10^{-4}$ M, $4\times10^{-4}$ M and $8\times10^{-4}$ M. These four samples share the same thickness of approximately 2.5 mm. Additionally, to further increase the optical density, we fixed the concentration at $8\times10^{-4}$ M and doubled the thickness to about 5 mm, as shown in Figure 6c. The selection of these concentrations and thicknesses was based on preliminary simulation results covering a wide range of parameters around optimum $\eta_{TRE}$. The absorption of the samples was measured using Ultraviolet–visible spectroscopy (UV-VIS, Shimadzu, UV-3600i Plus UV-VIS-NIR Spectrophotometer), as shown in Figure 6a. Based on Lambert-Beer's law, the MFP was calculated from the measured absorption, as shown in Figure 6b. It is evident that as the concentration increases, absorption increases while MFP decreases. There is a notable valley between 500 and 600 nm in the MFP, indicating that green photons are strongly absorbed by the device ($A_{ex}(\overline{red})$). It is also notable that the MFP in the 600-700 nm range for all samples is relatively high, indicating a comparatively low absorption ($A_{ex}(red)$) for red photons. Simulated and experimental results are illustrated in Figure 6d. As the fluorophore concentration (and hence optical density) increases, $\eta_{TRE}$ initially rises and then declines, peaking at a concentration of $4\times10^{-4}$ M and a thickness of 2.5 mm. The existence of this peak confirms the hypothesis based on the analysis of Equation (18). The Monte Carlo simulations align well with the experimental results obtained using a solar simulator (G2V Pico® Class AAA LED Solar Simulator) – details in Figure S7.

**3.3. Absorption spectrum**

Absorption spectrum is another crucial factor affecting HLSC efficiency. According to

Equation (18), increasing $A_{\text{ex}}(\overline{\text{red}})$ would undoubtedly improve $\eta_{\text{TRE}}$. However, increasing $A_{\text{ex}}(\text{red})$ would enhance the first term of $\eta_{\text{TRE}}$ but decrease the second. Based on this analysis, it is always beneficial for the absorption spectrum of fluorophores to cover a wide range of wavelengths shorter than red wavelengths. According to the discussion around Figure 2b, the intensity of solar spectrum is particularly strong between 400 and 600 nm, which is the most optimal range for fluorophores to absorb. The impact of $A_{\text{ex}}(\text{red})$ on $\eta_{\text{TRE}}$ can be expressed as:

$$(QY \times E_s - T_b) \times A_{\text{ex}}(\text{red}). \quad (20)$$

Based on this equation, absorption in the red range can only contribute positively to $\eta_{\text{TRE}}$ when both the QY and escape efficiency are high, and the transmittance is low. Therefore, determining whether the optimal absorption spectrum should cover the red range requires a more comprehensive analysis that accounts for all relevant factors.

To demonstrate the effect of the absorption spectrum on $\eta_{\text{TRE}}$, several simulations were conducted. To control the variables, all parameters were kept constant except for the absorption spectrum which was shifted within a range of 120 nm. In this paper, we assume negative values to indicate blue-shift and positive values to indicate red-shift. This adjustment ensured that the absorption of fluorophores covered a large range from 400 to 650 nm, as illustrated in Figure 7a.

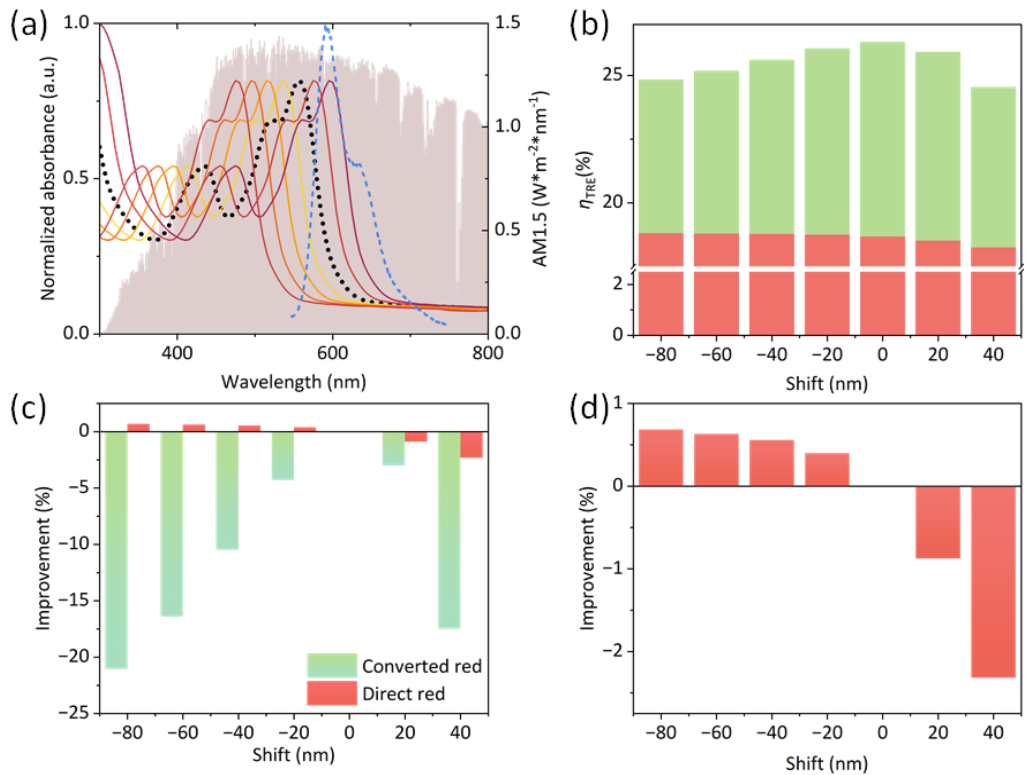

**Figure 7.** a) Normalized absorption spectra. The dotted black line and dashed blue line represent the real absorption spectrum and emission spectrum of Lumogen Red, respectively. The remaining lines are shifted absorption spectra. The shaded background corresponds to the AM 1.5 solar spectrum. b) Simulated $\eta_{TRE}$ of Lumogen Red in PDMS for all different absorption spectra. c) Improvement of $\eta_{DRE}$ and $\eta_{CRE}$ compared to real Lumogen Red absorption spectrum. d) Detailed improvement of $\eta_{DRE}$.

Simulated results shown in Figure 7b reveal how $\eta_{TRE}$ varies with shifting of the absorption spectra. As absorption spectra shift towards the red, from -80 nm to +40 nm, $\eta_{TRE}$ initially increases and then decreases. The highest $\eta_{TRE}$ is achieved with no red shift (0 nm), peaking at approximately 26.3%. The improvements in $\eta_{CRE}$ and $\eta_{DRE}$ compared to the 0-nm shift are plotted to analyse the effect of the absorption spectra, as shown in Figure 7c. The results indicate that shifting the absorption spectra primarily affects $\eta_{CRE}$ more than $\eta_{DRE}$. By blue-shifting the absorption spectra by 80 nm, for example, $\eta_{CRE}$ is reduced by over 20%. This dramatic drop is mainly attributed to the shifted absorption spectrum failing to cover the most intense part of the solar spectrum, similar to the condition observed with a +40 nm red-shifting. However, the effect of absorption spectra on $\eta_{DRE}$ is quite different, as shown in Figure 7d. $\eta_{DRE}$ drops significantly with shifts from -80 to 40 nm. This drop is mainly due to the substantial increase in the absorption of red photons. This result indicates that the optimum absorption spectrum should cover the highest energy regions (400-600 nm) to ensure sufficient and effective energy transfer.

**3.4. Emission spectrum**

An optimal absorption spectrum helps promote the absorption of undesirable colour energy. Whether this collected energy can be utilized by plants is determined by the emission spectrum of fluorophores. As shown in Figure 8a, the normalized absorption spectra of chlorophyll a and chlorophyll b peak in the blue and red regions. Most photoconversion techniques for horticultural applications primarily focus on red conversion for two main reasons. First, since most of the absorbed energy is in the green range, achieving up-conversion is challenging[21]. On the other hand, many commercial fluorophores can easily convert green light to red light[31,44,45]. Secondly, as shown in Figure 1, blue light can have side effects on plant growth, such as reducing leaf area. Based on this analysis, the most preferable emission spectrum should peak between 600 and 700 nm.

The impact of the emission spectrum on $\eta_{TRE}$ was verified through a series of simulations. To maintain controlled conditions, all parameters were held constant except for an 80 nm shift in the emission spectrum. This adjustment ensured that the fluorophores' emission spectrum ranged from 500 to 700 nm, as depicted in Figure 8a.

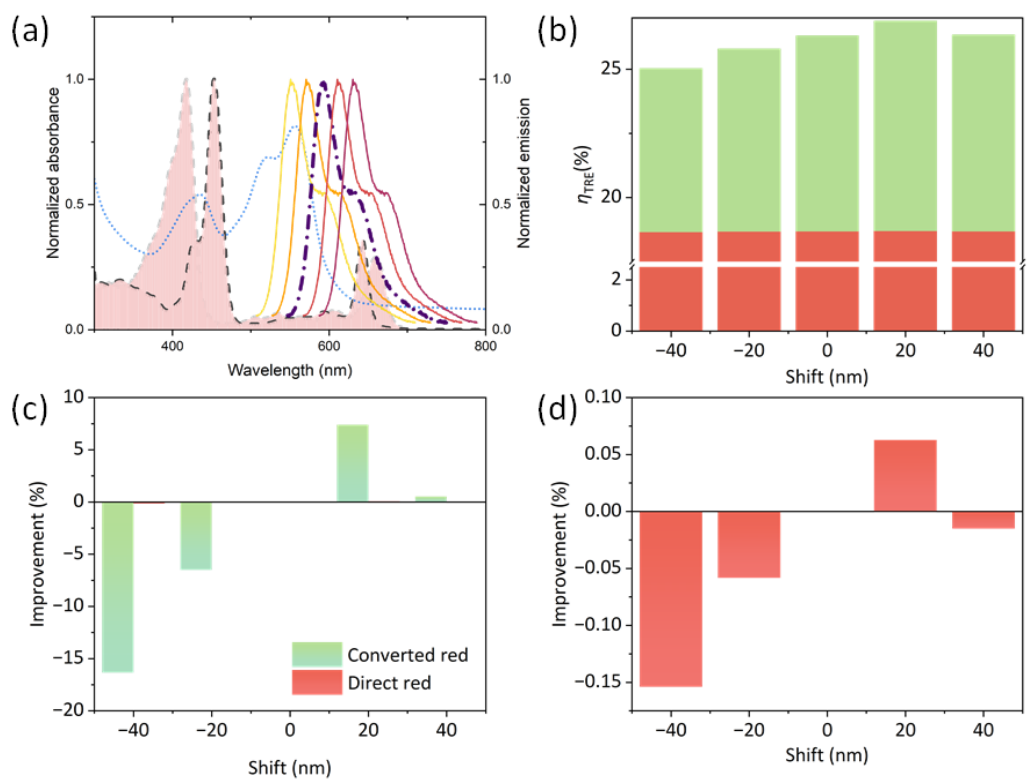

**Figure 8.** a) Normalized emission spectra. The dotted blue line and dot-dash line represent the real absorption spectrum and emission spectrum of Lumogen Red, respectively. The remaining lines are shifted emission spectra. The pink areas represent the normalized absorbance spectra of chlorophyll a and chlorophyll b. b) Simulated $\eta_{TRE}$ of Lumogen Red with different emission spectra. c) Improvement of $\eta_{DRE}$ and $\eta_{CRE}$ compared to real Lumogen Red. d) Detailed improvement of $\eta_{DRE}$.

Simulated results, depicted in Figure 8b, demonstrate how $\eta_{TRE}$ changes as emission spectra shift. As the emission spectra move towards the red, $\eta_{TRE}$ initially rises and then falls. The maximum $\eta_{TRE}$, approximately 26.9%, is observed for a 20 nm red shift. Figure 8c shows the changes in $\eta_{CRE}$ and $\eta_{DRE}$ in comparison to the 0-nm shift to analyse the impact of the emission spectra. A blue-shift of 40 nm in the emission spectra can decrease $\eta_{CRE}$ by more than 15%. This considerable decline is primarily because the shifted emission spectrum moves outside the optimal 600 to 700 nm range. Conversely, a 20 nm red shift centres the emission spectrum within the red range, resulting in the highest improvement of about 7.5%. However, further red-shifting reduces this benefit. The impact on $\eta_{DRE}$ mirrors that on $\eta_{CRE}$, as seen in Figure 8d, due to both improvements stemming from the re-emitted red photons. This result indicates that the optimum emission spectrum should align with the most desirable wavelengths for plant growth while avoiding overlap with the absorption spectrum.

### 3.5. Refractive index/QY

The impacts of refractive index and QY on $\eta_{TRE}$ are quite straightforward. According to Equation (18), decreasing the refractive index of the host material and increasing the QY both positively affect $\eta_{TRE}$. Several simulations were conducted to demonstrate the impact of these two parameters.

In fact, red emission trapping originates from the refractive index contrast. Ideally, the refractive index of the host material should equal that of the surrounding environment (usually air n~1). Figure 9a illustrates the variation in $\eta_{TRE}$ with the increase of the host material's refractive index. Interestingly, the highest $\eta_{TRE}$ is not achieved when the index is 1 but rather when it is 1.1. Beyond this point, $\eta_{TRE}$ continuously decreases with the increase in the refractive index, as expected. To analyse this phenomenon, $\eta_{CRE}$ and $\eta_{DRE}$ were simulated. As shown in Figure 9b, $\eta_{CRE}$ continuously decreases with increasing refractive index. However, $\eta_{DRE}$ shows a different trend, peaking at indices of 1.1 and 1.2. The higher $\eta_{DRE}$ at these indices contributes to the overall improvement in $\eta_{TRE}$ compared to when the index is 1.

This result shows that a slight refractive index contrast can enhance $\eta_{CRE}$. This enhancement is due to back-reflection on the top surface. Without any index contrast, scattered or re-emitted red photons by the fluorophores would be lost through the top surface. However, a slight index contrast between the top surface and the surrounding environment can back-reflect some of these photons, giving them another opportunity to escape from the bottom surface. This balance ultimately leads to the highest $\eta_{TRE}$ when the refractive index is 1.1.

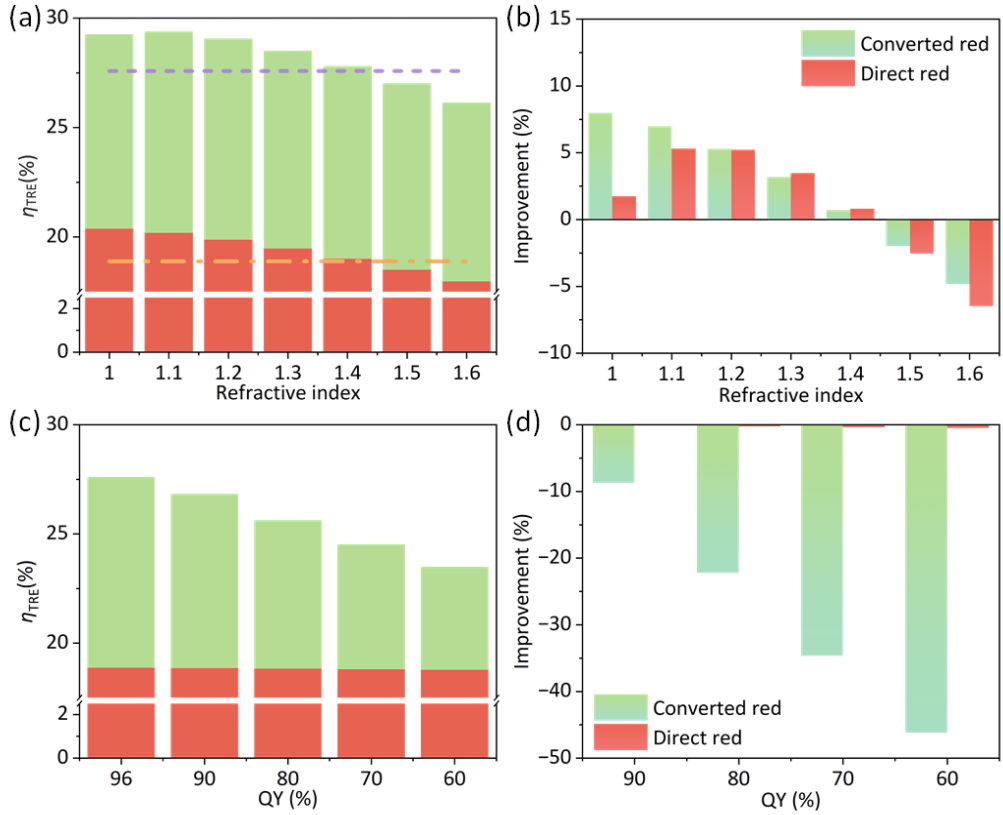

**Figure 9.** a) Simulated $\eta_{TRE}$ under different host indices. b) Improvement of $\eta_{DRE}$ and $\eta_{CRE}$ compared to a real PDMS host. The dashed purple line represents the $\eta_{TRE}$ of real PDMS, while the dashed-dot line represents the $\eta_{CRE}$ of real PDMS. c) Simulated $\eta_{TRE}$ with different QY. d) Improvement of $\eta_{DRE}$ and $\eta_{CRE}$ compared to real Lumogen Red QY.

Conversely, QY affects $\eta_{TRE}$ in a more straightforward manner, where a higher QY results in a higher $\eta_{TRE}$, as shown in Figure 9c. The improvements in $\eta_{CRE}$ and $\eta_{DRE}$, illustrated in Figure 9d, indicate that this enhancement primarily stems from the increase in $\eta_{CRE}$. When the QY is 60%, $\eta_{CRE}$ decreases by over 40% compared to when the QY is 96%. In contrast, $\eta_{DRE}$ remains almost unchanged with variations in QY, as depicted in Figure 9d. To further explore the impact of QY, a sample was subjected to a UV resistance test within a controlled chamber environment. After 146 days of UV exposure, both QY and absorption diminished, resulting in a significant reduction in the $\eta_{TRE}$ of the HLSC sample, as depicted in Figure S8.

**4. Discussion and Outlook**

**4.1. Guidelines for novel fluorescent and host materials**

In this study, we aim to provide guidelines for the development of novel fluorescent and host materials.

In the context of host materials for HLSCs, the real refractive index and extinction coefficient are crucial factors influencing performance. Our findings suggest that while host absorption is generally smaller in HLSCs, excessively high extinction coefficient within the 300-800 nm range can still adversely affect performance. Preferably, materials with a low real refractive index (ideally n~1.1) and low extinction coefficient should be used.

Polymers containing fluorine or those treated with fluorine often exhibit low real refractive indices[46]. This is attributed to the low polarizability of fluorine atoms, which renders the material less responsive to electromagnetic field. Fluorine's strong electronegativity and dense electron cloud contribute to the stability of chemical bonds, thereby reducing the material's ability to alter light's path. For instance, Poly(hexafluoropropylene oxide), commonly abbreviated as PHFPO, is reported to have a refractive index of 1.301[47], which is the lowest known to us. However, it is important to note that the use of Per- and Polyfluoroalkyl Substances (PFAS) has raised significant environmental concerns. Alternatively, introducing mesoporous structures within current host materials could significantly reduce the refractive index, potentially creating a medium with an index close to 1[27]. However, care has to be taken to avoid introducing unwanted scattering. Potentially, moth-eye nanostructures could be incorporated to create a gradient index, which would better guide photons to exit the device from the bottom surface[48,49].

In the development of novel fluorescent materials, such as new quantum dots, the QY of these materials is a critical consideration. High QY is essential not only for superior spectral conversion but also for minimizing loss channels. As discussed earlier, even minor reductions in QY can significantly compromise the performance of HLSC devices. Therefore, it is generally recommended that a QY of at least 80% is maintained for HLSC applications.

In the context of high QY, the absorption and emission spectra of fluorescent materials should also be better matched to the needs of plant growth. Generally, an ideal fluorescent material should have an absorption spectrum that is as broad as possible, encompassing wavelengths shorter than red light. Achieving this goal may involve exploring new types of quantum dots, such as investigating novel compositions and structures of quantum dots or utilizing host-guest composite luminescent systems[50]. This approach enables the transfer of a greater amount of solar energy to spectral regions preferred by plants, even under conditions where Stokes shifts are limited. Typically, the emission spectrum of fluorescent materials should include red and far-red parts. However, the distribution of the emission spectrum can be flexibly adjusted to cater to the diverse preferences of different plants, which is a key advantage of HLSCs in enhancing crop yield. This flexibility allows for the emission spectrum to be tailored to suit various crop preferences. As for the overlap between absorption and emission spectra, the ideal fluorescent material should ideally eliminate any overlap completely to minimize reabsorption. In the HLSC, this is more challenging because red and green wavelengths are adjacent to each other, making it likely that some overlap will always occur. Additionally, newly developed fluorophores with up-conversion capabilities are expected to convert energy not only from the green to blue region but also from the near-infrared to red region[20].

Regarding the optimal optical density, this represents a more complex issue. It is evident from our previous discussions that an HLSC system, with predetermined fluorescent and host materials, possesses an optimal optical density, which is determined by the optical properties of the fluorescent and host materials. The Monte Carlo ray tracing method is an effective approach for determining this optimal optical concentration under these conditions.

### 4.2. DRE and CRE

According to Equation (6), $\eta_{\text{TRE}}$ depends on both $\eta_{\text{DRE}}$ and $\eta_{\text{CRE}}$. Currently, it remains a challenge to simultaneously enhance $\eta_{\text{DRE}}$ and $\eta_{\text{CRE}}$ using existing light extraction techniques. This issue is a significant obstacle in achieving high-efficiency HLSCs compared to other devices also requiring light extraction, such as OLEDs, where no direct component is present. Present light extraction methods primarily aim to enhance $\eta_{\text{CRE}}$ rather than $\eta_{\text{DRE}}$, and many of these approaches adversely affect $\eta_{\text{DRE}}$. Given this trade-off, it is valuable to explore how much improvement in $\eta_{CRE}$ might offset a reduction in $\eta_{DRE}$, leading to an overall enhancement of $\eta_{TRE}$.

Variations in $\eta_{TRE}$ can be calculated using Equation (6):

$$\Delta\eta_{TRE} = \Delta\eta_{DRE} \times \rho_d + \Delta\eta_{CRE} \times \rho_c. \quad (21)$$

To ensure that $\Delta\eta_{TRE} > 0$, the relationship between $\Delta\eta_{DRE}$ and $\Delta\eta_{CRE}$ should be:

$$\frac{\Delta\eta_{DRE}}{\Delta\eta_{CRE}} > -\frac{\rho_c}{\rho_d}. \quad (22)$$

According to Equation (22), as the percentage of red components in the spectrum increases, it becomes more challenging to develop an efficient HLSC. This equation establishes criteria for adapting HLSCs to the varying solar spectra in different regions of the world.

### 4.3. Surface patterning

Surface patterning primarily influences $\eta_{TRE}$ by altering $E_s$ and $T_b$ in Equation (18). It is evident that increasing both $E_s$ and $T_b$ benefits $\eta_{TRE}$. Patterning the bottom surface affect both $E_s$ and $T_b$, while patterning the top affects mostly $E_s$.

An optimized surface pattern on the bottom surface should be capable of extracting incident rays from a broad range of directions, considering that the converted photons are re-emitted isotropically. Additionally, this pattern should not impede the direct red photons hitting the bottom surface. Similarly, the pattern on the top surface should be designed to back-reflect photons, recycling those that tend to escape from the top surface of the device. However, it is important to note that a patterned top surface can also increase Fresnel reflection, which could ultimately reduce the number of photons entering the device in the first place.

Surface patterning is the most promising method to improve $\eta_{TRE}$ in HLSCs. Increasing escape efficiency without affecting direct red emission is quite challenging. Microstructures such as micro-cone, micro-lens, and micro-pyramid arrays are expected to be utilized to achieve this goal[28,36,37]. The designs of the top and bottom surfaces should be conducted in tandem because top and bottom surfaces interact with one another influencing $E_s$ and $T_b$.

In addition to light extraction structures, a Distributed Bragg Reflector (DBR) is also a potential tool for improving the overall performance of HLSCs. A DBR can reflect photons within a specific wavelength range while transmitting the rest[38]. A well-designed DBR, patterned on the bottom surface of the HLSC, would transmit only the wavelengths preferred by plants (600-700 nm) while reflecting the rest. This configuration would not only allow high transmittance for red photons but also back-reflect undesirable photons for reabsorption, thereby increasing absorbance and $\eta_{CRE}$, ultimately enhancing the $\eta_{TRE}$.

5. Conclusion

In conclusion, HLSCs represent a recent innovative advancement, building upon the traditional LSCs. This study addresses the specific requirements of horticulture to clarify the motivation for employing HLSCs. We have re-evaluated and redefined the optical metrics of HLSCs for the first time, analysing potential loss channels for both direct red emission and converted emission. Through the use of the Monte Carlo ray tracing method and experimental data, we have explored the factors influencing these loss channels. Our work offers a foundational discussion on HLSCs and provides a design guideline for future research in this area.